\documentclass[twocolumn,showpacs,preprintnumbers,amsmath,amssymb]{revtex4}

\usepackage{graphicx}
\usepackage{dcolumn}
\usepackage{bm}

\begin{document}

\title{Saddle point states in 2D superconducting film biased near the depairing current}

\author{D.Y. Vodolazov}
\email{vodolazov@ipmras.ru} \affiliation{Institute for Physics of
Microstructures, Russian Academy of Sciences, 603950, Nizhny
Novgorod, GSP-105, Russia}

\date{\today}


\begin{abstract}

The structure and energy of saddle point (SP) states in 2D
superconducting film of finite width $w$ with transport current
$I$ are found in the framework of Ginzburg-Landau model. We show
that very near depairing current $I_{dep}$ the SP state with a
vortex does not exist and it transforms to 2D nucleus state, which
is a finite region with partially suppressed order parameter. It
is also shown that for slightly lower currents the contribution of
the vortex core energy is important for SP state with a vortex and
it cannot be neglected for $I\gtrsim 0.6 I_{dep}$. It is
demonstrated that in the film with local current concentration
near the bend the energy of SP state may be much smaller than one
in the straight film and it favors the effect of fluctuations in
such a samples.

\end{abstract}

\maketitle

\section{Introduction}

Many physical systems have several metastable states at fixed
external parameters (temperature, magnetic field, etc.) which
correspond to different local minima of their free energy. As an
example it could be mentioned the different configuration of DNA
molecule \cite{Daniels}, vortex 'molecules' in the mesoscopic
superconductors \cite{Baelus,Misko}, or magnetic states of the
lattice of magnetic nanocaps \cite{Sapozhnikov}. The metastable
states are usually separated from each other by the energy barrier
$\Delta F$ which could be overcome due to thermal activation. If
the height of the energy barrier is much larger than the thermal
energy $k_B T$ then such a transitions are relatively rare events
and one may estimate their rate as $\sim exp(-\Delta F/k_B T)$
\cite{Hanggi}. The standard procedure to calculate $\Delta F$ is
to find the saddle point states which correspond to the local
maxima of the free energy and find the energy difference between
them and the metastable states \cite{Daniels,Baelus,Misko,Hanggi}.

In the superconductors with transport current less than critical
one this idealogy could be also applied for calculation of the
finite resistance appearing due to fluctuations. For 1D
superconducting wires Langer and Ambegaokar (LA)  \cite{Langer} in
framework of the Ginzburg-Landau (GL) model find that in the
saddle point state the superconducting order parameter
$\psi=|\psi|e^{i \varphi}$ is partially suppressed in the finite
region along a wire and the amplitude of suppression depends on
the current. If one starts from such a 1D nucleus state the time
evolution of the order parameter will inevitably lead to the phase
slip \cite{Langer} and to the finite resistance $R_{wire}$. LA
find dependence of $\Delta F$ both on the current and temperature
which with good accuracy \cite{Tinkham1} could be written as
\begin{equation}
\frac{\Delta F_{LA}}{F_0}=\frac{2\sqrt{2}}{3\pi} \frac{w}{\xi}
\left (1-\frac{I}{I_{dep}}\right)^{5/4}
\end{equation}
where $F_0=\Phi_0^2d/16\pi^2\lambda^2$, $\Phi_0$ is a magnetic
flux quantum, $w$ is a width and $d$ is a thickness of the wire
respectively, $\lambda$ is a  London penetration depth, $\xi$ is a
coherence length and
$I_{dep}=c\Phi_0wd/12\sqrt{3}\pi^2\xi\lambda^2$ is a depairing
current in the Ginzburg-Landau model. According to the general
concept $R_{wire}= \nu \, exp(-\Delta F_{LA}/k_BT)$ where the
pre-exponential factor $\nu$ is calculated in Refs.
\cite{McCumber,Golubev}.

In contrast to 1D wire in thin ($d \ll \lambda$) 2D film several
saddle point states exist at given value of the current. Further
we discuss the case of the relatively narrow film with $\xi \ll
w<\lambda^2/d$ (this condition ensures the uniformity of the
current distribution over the width of the superconductor with
transport current in the ground state). At the present time it is
distinguished three types of the saddle point states in such a
samples: i) the state with a single vortex
\cite{Kogan,Maksimova,Clem,Qiu1,Bartolf,Bulaevskii}, ii) the
vortex-antivortex (VA) state \cite{Bartolf,Bulaevskii,Mooij,Qiu3}
and iii) LA state extended to 2D case
\cite{Qiu1,Bartolf,Bulaevskii}. In Ref. \cite{Bulaevskii} it was
argued that VA state has at least twice larger energy than the
single vortex state and LA-like state is the most energetically
unfavorable among the considered states at all currents $I \leq
I_{dep}$. However in Ref. \cite{Bulaevskii} the current dependence
of $\Delta F_{LA}$ (see Eq. (1)) was not taken into account. If
one takes the result for the energy of the single vortex (V) state
near depairing current $1-I/I_{dep} \ll 1 $ (found in the London
model \cite{Qiu1,Bartolf,Bulaevskii})
\begin{equation}
\frac{\Delta F_V}{F_0}\simeq 1-\frac{I}{I_{dep}}
\end{equation}
and compare it with Eq. (1) then it is easy to see that even for
wide films $w\gg \xi$ there is a finite region of the currents
$1-(3\pi \xi/2\sqrt{2}w)^4<I/I_{dep} \leq 1$ where $\Delta
F_{LA}<\Delta F_V$. But such a quantitative comparison is not
correct because Eq. (2) does not contain the energy of the vortex
core ($E_{core}\sim 0.38 F_0$ for vortex located far from the
edges \cite{Steijic}) and, as we show below, it seriously
underestimates $\Delta F_V$ at $I \gtrsim 0.6 I_{dep}$.

The effect of the vortex core was taken into account in the recent
work \cite{Qiu1} using GL model. But authors focused on the region
of small currents $I\ll I_{dep}$ (where contribution of $E_{core}$
to $\Delta F_V$ is relatively small) and they concluded that for
films with $w>w_c\simeq 4.4 \xi$ the energy of single vortex state
is lower than the energy of LA state. Below we show that vortex SP
state does not exist at $I\sim I_{dep}$ and it transforms to {\it
vortex free} 2D nucleus (which is a finite region with partially
suppressed order parameter) located near the edge of the film. We
confirm our above estimation that very near $I_{dep}$ even for
wide films $w \gg w_c$ saddle point state of Langer and Ambegaokar
has a lowest energy among over SP states. We also study
practically important case of the film with one $180^0$ bend which
models the part of the superconducting meander used in
superconducting single photon detectors
\cite{Gol'tsman,Divochiy,Marsili}. We show that due to current
concentration near the bend the jump to the saddle point state
needs much less energy at $I\to I_c$ in comparison with 2D film
with uniform current distribution, which promotes the effect of
fluctuations in such a samples. This result is rather general and
could be applied to any superconducting system with a local
current concentration.

\section{Method}

To find the state which is a solution of {\it stationary}
Ginzburg-Landau equations but is {\it unstable} one (saddle point
state) we use the following numerical procedure. In case when we
search for SP state {\it with a vortex} we first put (as an
initial condition) the phase distribution corresponding to the
vortex seating in the point (n,m) of the discrete grid and
additionally fix the phase difference between adjacent points
$\varphi(n,m+1)-\varphi(n,m-1)=\pi$ (which, in some respect, pins
the vortex at the point (n,m)) at any time step. The solution of
the stationary GL equations we find by using the relaxation method
(by adding the time derivative $\partial \psi/\partial t$ in GL
equations and waiting when it goes to zero). By variation of the
current (as an external parameter) it is possible to find a
stationary state when the vortex does not move from the point
(n,m). At low currents there are several points where the vortex
position could be fixed by this method for given $I$. By our
definition the one with the highest energy corresponds to SP state
(it corresponds to the local maximum of Gibbs energy as function
of the vortex position in the London approach
\cite{Kogan,Maksimova,Clem,Bartolf,Bulaevskii}). After reaching
such a state the energy difference could be found using the
following expression
\begin{equation}
\Delta F=F_{saddle}-F_{ground}-\frac{\hbar}{2e}I\Delta\varphi
\end{equation}
where $\Delta \varphi$ is an additional phase difference between
ends of the film which appears in SP state in comparison with the
ground state and $F_{saddle}$ and $F_{ground}$ are the
Ginzburg-Landau free energies of the saddle point and ground
states respectively.

To find the {\it vortex free} saddle point state we fix the
magnitude of the order parameter $|\psi|>0$ in one point at the
edge and allow to vary $\psi$ in all other points. Than we
increase the current up to the moment when such a state becomes
nonstationary. By our definition we find the vortex free SP state
corresponding to {\it the given value of the current}. We checked
that if we start from this state and let $\psi$ vary everywhere in
the film, the vortex is nucleated in the point where we initially
fix $|\psi|$ and passes across the film. This finding is an
extension to 2D case of the main idea of LA that if one starts
from SP state with finite $|\psi|$ everywhere in the sample the
time evolution of the order parameter will inevitably lead to the
phase slip and voltage pulse.

The proposed method is much simpler (from point of view numerical
procedure) than the methods used in Refs.
\cite{Schweigert,Pogosov} for finding SP states in the mesoscopic
superconducting disk in magnetic field or 2D film with a transport
current \cite{Qiu1,Qiu3}. Similar procedure to fix the vortex
position was used, for example, in Ref. \cite{Priour}. Moreover
our method could be easily applied to 2D samples of arbitrary
geometry (triangles, disks, stars, etc.) and to 3D case to find
the {\it vortex free} saddle point states. We checked that the
method gives the same results for 1D saddle point state (both for
the spatial dependence of order parameter and excess energy
$\Delta F$) as analytically found in Ref. \cite{Langer}. Validity
of the proposed method is also supported by our results for 2D
film because they coincide with results found in the London model
at low currents.

In numerical calculations the step of the rectangular grid was
$\delta x=\delta y=\xi/4$ and width of the film varied from 4.5 up
to 30 $\xi$. The length of the film was chosen $L=4 w$ (which is
long enough to neglect the effect of finite length). The boundary
conditions, written here in dimensionless units (for units see for
example \cite{Pogosov}), were the following: $\psi^*\nabla
\psi|_{y=\pm L/2}=iI/wd$ and $\nabla \psi|_{x=\pm w/2}=0$.

\section{Results}

In Fig. 1 the dependence of $\Delta F$ on the current for
different SP states is present for the film with $w=4.5 \xi$. We
should note that vortex in such a narrow film has strongly
deformed core (see upper inset in Fig. 1) and it resembles more a
Josephson vortex than an Abrikosov one \cite{Likharev}. In wider
films (see Figs. 2,3) the deformation of a vortex core occurs when
the vortex seats near the edge on the distance $\Delta x \lesssim
2 \xi$ (see insets in Figs. 2,3). Similar result was found for the
vortex placed near the artificial defect (see Fig. 2 in Ref.
\cite{Priour}) and the edge of the superconducting disk (see Fig.
3 in Ref. \cite{Schweigert}) or film (see Fig. 2 in Ref.
\cite{Qiu1}).
\begin{figure}[hbtp]
\includegraphics[width=0.53\textwidth]{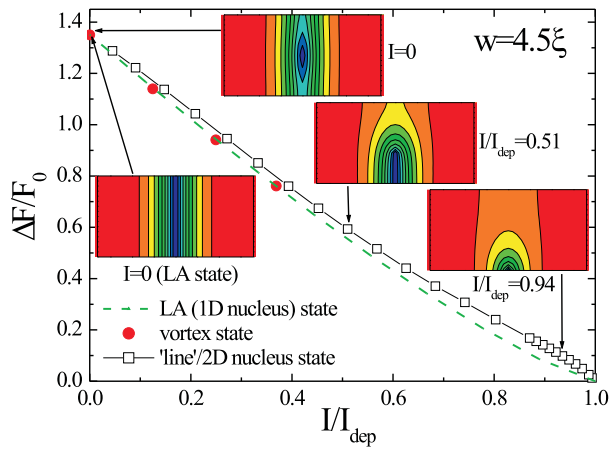}
\caption{Energy of saddle point states of three kinds: LA (green
dashed curve), vortex/'line' (red circles/rare empty squares) and
2D nucleus (dense empty squares) in the film with $w=4.5 \xi$. In
the insets we present the contour plot of $|\psi|$ at different
currents and different kinds of SP states.}
\end{figure}

Unfortunately our numerical method does not allow to find SP state
with a vortex when $\Delta x < 1.5 \xi$ (last red circle at the
largest current in Figs. (1-3) corresponds to $\Delta x = 1.5
\xi$) because we were not able to find the stationary solution of
Ginzburg-Landau equation with additional condition
$\varphi(n,m+1)-\varphi(n,m-1)=\pi$ (this effective pinning
'force' becomes not sufficiently strong to pin the vortex very
near the edge). But we notice that if we fix $|\psi|=0$ along
finite length line near the edge (see inset in Fig. 1 where
contour plot of $|\psi|$ at $I/I_{dep}=0.51$ is shown) and find
the stationary solution of GL equations with this additional
condition then the excess energy $\Delta F$ of such a 'line' state
(see rare empty squares in Figs. 1-3) is close to the energy of
the vortex state (when vortex seats relatively close to the edge -
see Figs. 2,3). Because of that one may approximate the energy of
the vortex state when $\Delta x < 1.5 \xi$ by the energy of the
'line' state with length $l<2 \xi$.
\begin{figure}[hbtp]
\includegraphics[width=0.53\textwidth]{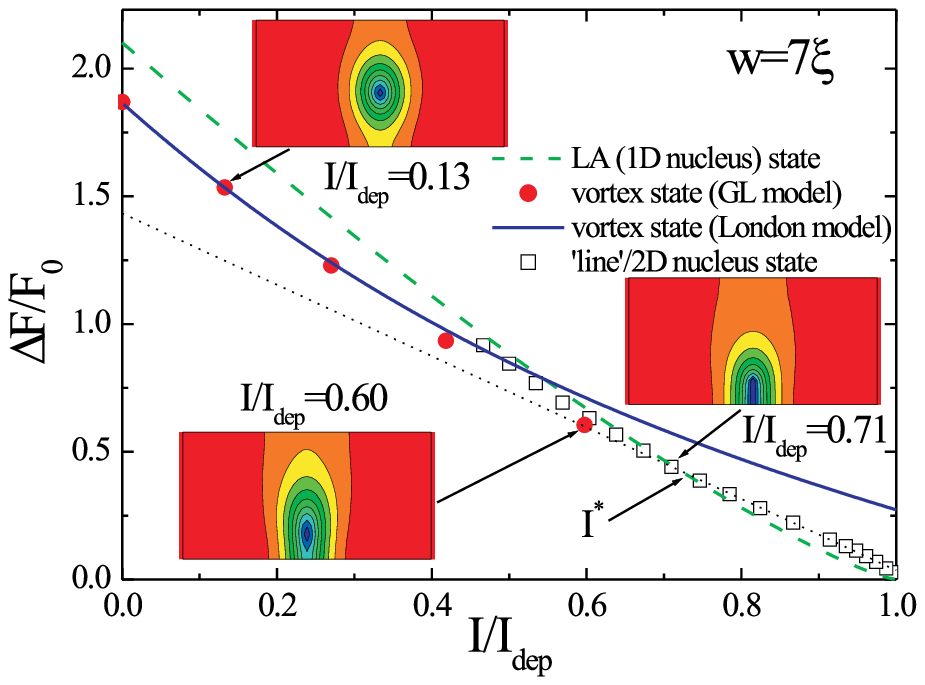}
\caption{Energy of saddle point states of three kinds: LA (green
dashed curve), vortex/'line' (red circles/rare empty squares) and
2D nucleus (dense empty squares) in the film with $w=7 \xi$. Blue
curve corresponds to Eq. (4). Black dotted line is the dependence
$\Delta F/F_0=1.43(1-I/1.026I_{dep})$ which fits well (deviation
less than $2\%$) our numerical results at $0.6 \leq I/I_{dep} \leq
0.97$. At $I>I^*\simeq 0.73 I_{dep}$ LA state has a lowest
energy.}
\end{figure}

At currents $I\sim I_{dep}$ the vortex('line') state transforms to
the vortex-free SP state (dense empty squares in Figs. 1-3) when
the phase circulation along any closed contour in the film is
equal to zero and $|\psi|>0$ everywhere in the film. To find it we
fix the amplitude of the order parameter $|\psi|$ in one point at
the edge. Because of proximity effect and that $I\sim I_{dep}$ it
leads to suppression of $|\psi|$ in the relatively large region
around this point (see inset in Fig. 1 at $I/I_{dep}=0.94$).
Further we call it as 2D nucleus state to distinguish it from 1D
nucleus state of LA (compare insets in Fig. 1 at $I=0$ and at
$I/I_{dep}=0.94$).

In Figs. (2,3) we also plot dependence of $\Delta F$ for the
energy of the vortex SP state found in the London limit
\cite{Qiu1,Bartolf,Bulaevskii} (solid blue curve)
\begin{eqnarray}
\frac{\Delta F_V}{F_0}=-\frac{1}{2}ln
\left(1+\frac{I^2}{\alpha^2I_{dep}^2}\right)- \frac{I}{\alpha
I_{dep}} \tan^{-1} \left[\frac{\alpha I_{dep}}{ I} \right] +
\nonumber
\\
+ \epsilon+ln\left(\frac{2w}{\pi \xi}\right)
\end{eqnarray}
where $\alpha=3\sqrt{3}\pi\xi/4w$ and we add the energy of the
vortex core $E_{core}=\epsilon F_0$. Numerical coefficient
$\epsilon$ is found from comparison of Eq. (4) with numerical
result at $I=0$ and it is present in Table 1 for different widths.
Notice the good agreement between GL and London model at
$I\lesssim 0.6 I_{dep}$. At larger currents vortex is located on
the distance $\Delta x \lesssim 2 \xi$ from the edge and one has
to take into account deformation of the core which provides the
dependence $\epsilon(I)$. It leads to large discrepancy between
London (with $\epsilon=const$) and GL models at $I \gtrsim 0.6
I_{dep}$ (see Figs. 2,3). Moreover at $I\sim I_{dep}$ vortex state
transforms to the 2D nucleus state which cannot be found in the
London limit.

\begin{table}
\caption{\label{tab:table1}Values of coefficients in the fitting
expressions for energy of vortex/2D nucleus saddle point states
(Eq. (5)) and vortex core energy when vortex seats in the center
of the film (coefficient $\epsilon$ in Eq. (4)).}
\begin{ruledtabular}
\begin{tabular}{cccccc}
$w/\xi$ &A &B  &C &n & $\epsilon$  \\
7& 1.43 & 1.026 & 0.89 & 0.7 & 0.37  \\
10& 1.67 & 1.026 & 1.02 &0.7 & 0.38  \\
15& 1.85 & 1.028 & 0.88 &0.6 & 0.38  \\
30& 1.88 & 1.034 & 0.68 &0.5 & 0.38  \\
\end{tabular}
\end{ruledtabular}
\end{table}

Our numerical results at $I/I_{dep}\gtrsim 0.6$ could be fitted
(examples of fitting are present in Fig. 2,3 and inset in Fig. 5)
by the following functions
\begin{equation}
 \frac{\Delta F}{F_0} \simeq \left \{ \begin{array}{ll}
 \displaystyle{A(1-I/BI_{dep})}, & 0.6 \leq I/I_{dep} \leq 0.97 ,\\
 \displaystyle{C(1-I/I_{dep})^n },& 0.95 \leq I/I_{dep} \leq 1  .
\end{array} \right.
\end{equation}
Coefficients A,B,C and power n for different widths are listed in
Table 1. Note that coefficient A is almost twice larger than the
result which follows from the London model (see Eq. (2)). It
reflects the contribution of $E_{core}(I)$ to $\Delta F$ which for
$I \gtrsim  0.6 I_{dep}$ cannot be neglected and $E_{core}$
gradually decreases with increasing $I$.
\begin{figure}[hbtp]
\includegraphics[width=0.53\textwidth]{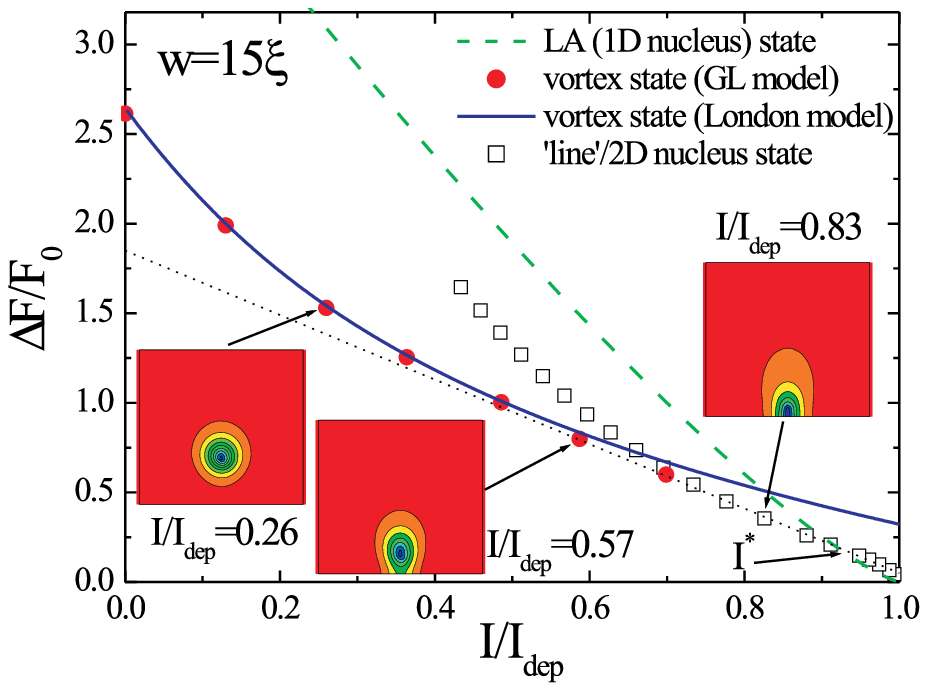}
\caption{Energy of saddle point states of three kinds: LA (green
dashed curve), vortex/'line' (red circles/rare empty squares) and
2D nucleus (dense empty squares near depairing current) in the
film with $w=15 \xi$. Blue curve corresponds to Eq. (4). Black
dotted line is the dependence $\Delta
F/F_0=1.85(1-I/1.028I_{dep})$ which fits well (deviation less than
$2\%$) our numerical results at $0.6\leq I/I_{dep} \leq 0.97$. At
$I>I^*\simeq 0.92 I_{dep}$ LA state has a lowest energy.}
\end{figure}

At $I/I_{dep} \sim 1$ the power $n<1$ (see Table 1) in Eq. (5)
tells one that there is a finite (but rather narrow for wide films
with $w \gg \xi$) interval of currents $I^*(w)<I<I_{dep}$ where 1D
nucleus (LA) state has the lowest energy (see Figs. 2,3). This
counterintuitive, at first sight, result is explained by the
presence of the last term in right hand side of Eq. (3). Although
in LA state the order parameter is suppressed over whole width
(see inset at $I=0$ in Fig. 1) and it costs more condensation
energy than in vortex/2D nucleus state, the phase difference
$\Delta \varphi$ is much larger in LA state than in other SP
states at $I\sim I_{dep}$ and it brings above result.
\begin{figure}[hbtp]
\includegraphics[width=0.53\textwidth]{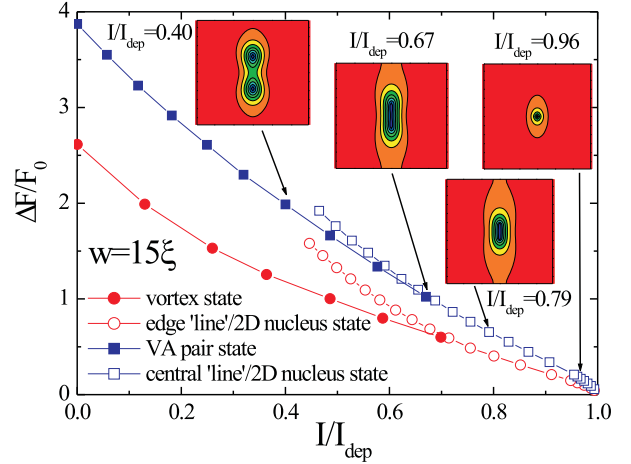}
\caption{Energy of saddle point states of different kinds: single
vortex (red circles), vortex-antivortex pair (blue squares) and
'line'/2D nucleus located at the edge (empty rare/dense circles)
and in the center (empty rare/dense squares) of the film.}
\end{figure}

Before we consider only single vortex state and 2D nucleus which
is located near the edge of the film. In Fig. 4 we demonstrate
that the energy of 2D nucleus located in the center of the film is
larger than the energy of the edge 2D nucleus. The
vortex/antivortex state which is transformed from 2D nucleus state
at lower currents (see solid and rare empty squares in Fig. 4) has
energy larger than the single vortex state. The similar result was
found in London limit where the difference reaches two times
between $\Delta F_V$ and $\Delta F_{VA}$ \cite{Bulaevskii,Qiu3}.

We also find the saddle point states in 2D superconducting film
with a $180^0$ bend - see left inset in Fig. 5. In our simulations
we choose the width of the film $w=10 \xi$, the length of the
sample in the bend region L=2w and two widths of the slit:
$w_{slit}=2.5 \xi$ (shown in the left inset in Fig. 5) and
$w_{slit}=10 \xi$. In Fig. 5 we present our results for the energy
of single vortex and edge 2D nucleus states. Note that the current
in Fig. 5 is normalized to the critical current of the sample and
not to the $I_{dep}$ as in Figs. 1-4 ($I_c=0.85 I_{dep}$ for film
with $w_{slit}=2.5 \xi,$ $I_c=0.91 I_{dep}$ for film with
$w_{slit}=10 \xi$ and $I_c=I_{dep}$ for straight film).

We want to stress here that at $I\sim I_c$ the energy of SP states
is considerably lower in the film with bend than without it
(compare results present in right inset in Fig. 5) taken at {\it
the same ratio} $I/I_c$. And the stronger the current
concentration near the bend (which is manifested in lower value of
$I_c$ for smaller $w_{slit}$) the smaller $\Delta F$. We expect
that the smallest $\Delta F$ could be reached in case of
infinitesimally narrow and long crack near the edge of the film
\cite{Aladyshkin,Vodolazov} which provides the maximal current
concentration and maximal suppression of $I_c$. We explain this
effect by the partial suppression of the order parameter (on the
scale of about several $\xi$) in the region with the strongest
current concentration even in the ground state. As a result it
takes less energy to jump to SP state from the ground state with
already locally suppressed $|\psi|$. The proof of this idea also
comes from our results at low currents where $\Delta F$ differs a
little for straight film and film with a bend (see Fig. 5). At low
currents the suppression of $|\psi|$ near the bend is weak and it
slightly influences $\Delta F$.

The decrease of $\Delta F$ in the 1D wire with constriction was
also found in Ref. \cite{Qiu2}. But authors of \cite{Qiu2} did not
present $\Delta F$ as function of $I/I_c$ and it is not clear is
their result connected with reduction of $I_c$ (which gives
trivial $\Delta F \to 0$ at current $I=I_c<I_{dep}$ and $\Delta
F<\Delta F_{LA}$ at $I<I_c$) or with a change in the power n in
asymptotics $\Delta F \sim (1-I/I_c)^n$ as in our case (see right
inset in Fig. 5).
\begin{figure}[hbtp]
\includegraphics[width=0.53\textwidth]{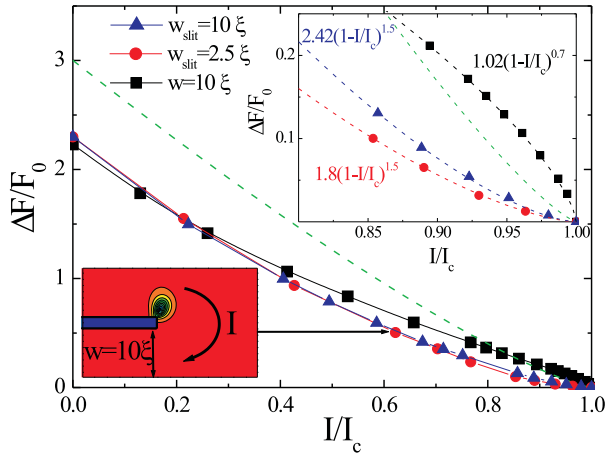}
\caption{Energy of saddle point states (vortex and 2D nucleus) for
the film with a bend and slits of different width: $w_{slit}=2.5
\xi$ (red circles) and $w_{slit}=10 \xi$ (blue triangles) and the
film without bend with $w=10 \xi$ (black squares). In the right
inset we present the zoom at $I\sim I_c$ which shows the energy of
2D nucleus SP states with dashed curves as the fitting of the
numerical results. Green dashed curve shows the energy of LA state
for straight film with $w=10 \xi$.}
\end{figure}

\section{Conclusion}

It is found that in 2D superconducting film with spatially uniform
current distribution in the ground state the lowest energy saddle
point state at $I\sim I_{dep}$ corresponds to the state of Langer
and Ambegaokar found for 1D wire. Only at $I<I^*$ (where
$I^*(w)<I_{dep}$ even for wide films $w\gg \xi$) the lowest energy
SP state corresponds either to edge 2D nucleus state or to state
with a vortex located next to the edge and having strongly
modified core. At currents $I \lesssim 0.6 I_{dep}$ (for films
with $w \gtrsim 7 \xi$) the lowest energy saddle point state
corresponds to the single vortex with ordinary core and the
results of the London model are recovered. We demonstrate that in
the film with current concentration, which leads to the local
spatial variation of the order parameter in the ground state, the
energy of 2D nucleus SP state may be much lower than the energy of
any saddle point states in the film with uniform current
distribution taken at the same ratio $I/I_c \sim 1$.

The last result has an important practical consequence. It shows
that if in the sample there are places with strong current
concentration (bends, geometrical defects of the edge and so on)
it favors the effect of fluctuations near $I_c$ because of much
low, in comparison with straight film without defects, energy
barrier. For example, if $k_BT=0.1 F_0$ the switching of the
straight film with $w=10 \xi$ to resistive state roughly occurs at
$I\simeq 0.96 I_c=0.96I_{dep}$ (at this current $\Delta F=0.1 F_0$
- see right inset in Fig. 5). But for the film with a bend and
$w_{slit}=2.5 \xi$ the energy barrier $\Delta F=0.1 F_0$ is
reached at $I\simeq 0.85 I_{c} \simeq 0.72 I_{dep}$ and transition
to resistive state occurs at much lower current than one could
expect from results for straight film.

We believe that the found results are valid not only near $T_c$
(where Ginzburg-Landau model is quantitatively valid) but at lower
temperatures too. Indeed, in the recent work \cite{Semenov}
asymptotics $\Delta F \sim (1-I/I_{dep})^{5/4}$ for $I\to I_{dep}$
was confirmed for LA like state in 'dirty' 1D superconducting wire
at any temperature. In the 'pure' 1D wire only $15\%$ difference
with LA result at $T \to 0$ was noticed in Ref. \cite{Zharov} in
the limit $I \to 0$ \cite{self1}.

\begin{acknowledgments}
This work was supported by the Russian Foundation for Basic
Research and Russian Agency of Education under the Federal Target
Programme "Scientific and educational personnel of innovative
Russia in 2009-2013".
\end{acknowledgments}


\begin{references}
\bibitem{self1} In Ref. \cite{Zharov} authors did not take into account
the additional term in the free energy, which describes the work
done by the external source of current $\sim I \Delta \varphi$
(see discussion in Ref. \cite{Semenov}) and they find the finite
$\Delta F$ even at $I\to I_{dep}$ (see inset in Fig. 1 in Ref.
\cite{Zharov}).
\end{references}
\end{document}